# Reversible photo-switching in a cell-sized vesicle

Tsutomu Hamada[†], Yuko T. Sato[†*], Takeshi Nagasaki[‡], and Kenichi Yoshikawa*[†]

*Department of Physics, Graduate School of Science, Kyoto University, Kyoto 606-8502, JAPAN*

*Department of Applied and Bioapplied Chemistry, Graduate School of Engineering, Osaka City University, Osaka 558-8585, JAPAN*

E-mail: hamada@chem.scphys.kyoto-u.ac.jp

In biological systems, change in the conformation of photosensitive molecules embedded in membranes, such as rhodopsin, phytochrome, etc., play important physiological roles.[1] Many studies have been conducted to prepare artificial photosensitive vesicles, which is also expected to be applied for drug delivery[2], and have reported changes in the permeability of ions and/or water-soluble compounds across the membrane upon photo-isomerization.[3] In these studies, small vesicles (~100 nm) have frequently been used, which implies that the direct observation of morphological change in individual vesicles is impossible. Unfortunately, these small vesicles are usually unstable due to their high curvature and undergo spontaneous changes such as fusion, breakdown, and aggregation even in the absence of external stimuli. In contrast, cell-sized vesicles ($\geq$ 10 μm) are rather stable and can be used as a suitable model system for observing transformational processes[4] and biochemical reactions inside them[5] in real-time. Along these lines, it has recently been found that transmembrane protein activity influences the membrane tension in a cell-sized vesicular system.[6]

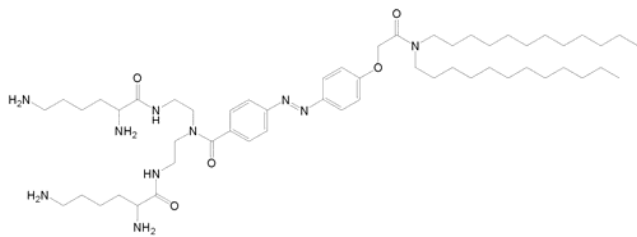

**Chart 1.** Chemical formula of KAON12.

In this study, we designed and synthesized a photosensitive amphiphilic molecule containing azobenzene (KAON12)[7]; the conformation (trans or cis) of this molecule can be switched by light (Chart 1). We prepared cell-sized vesicles with this photosensitive molecule together with phospholipid, and observed changes in its morphology by phase-contrast microscopy. Figure 1A shows the effect of photo-irradiation on a spherical vesicle. To clarify the difference in fluctuation behavior between isomers, changes in the radius and its distribution were measured (Fig. 1C), where we defined the radius along the contour $r(\theta)$ and the angle $\theta$ as in Fig. 1B. Much greater fluctuation is observed after UV treatment than after VIS (Visible light) treatment, indicating that a vesicle with a cis-azo conformation exhibits surplus membrane area to encapsulate the inner aqueous solution. In contrast, the small fluctuation on the periphery of the vesicle suggests that the membrane area just fits the spherical surface of the inner liquid phase. Figure 2A shows the results of the photo-irradiation on an asymmetrical vesicle, which is spontaneously formed through natural swelling of the lipid film. After UV irradiation, the vesicle exhibits budding. Interestingly, the budded vesicle transforms back to the original ellipsoidal shape upon treatment with green light. This reversible change in morphology is observed more than ten times (Fig. 2B). While the pathway of the transformation varies somewhat between the forward and reverse processes, the switching between the two stable states is reversible. Thus, under UV irradiation, an asymmetric deformed vesicle undergoes a budding transition, whereas a symmetrical sphere vesicle exhibits enhanced fluctuation without any significant morphological change.

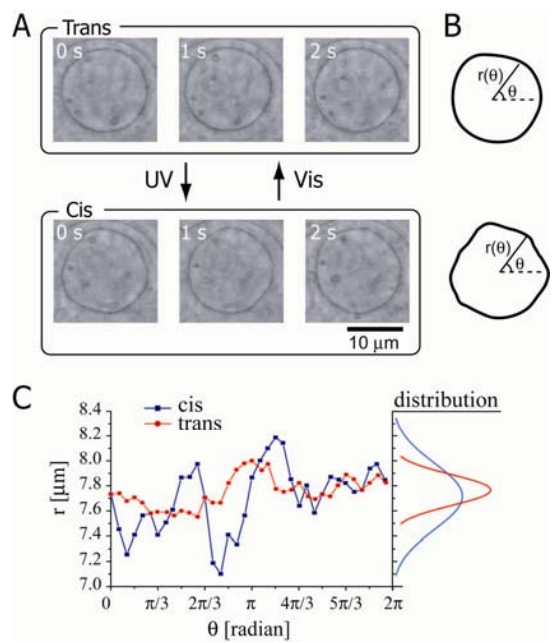

**Figure 1.** A: Phase-contrast images of the difference in the fluctuation behavior between trans and cis-azo vesicles at 1-second intervals. B: Schematic representation of the difference in membrane fluctuation. C: Spatial fluctuation in the radius around the contour together with its distribution profile.

We measured Π-A curves of a langmuir monolayer with a trans-azo or cis-azo conformation in the same concentration ratio as in the microscopic observation [see Supporting Information]. The surface area per molecule in the cis-azo form is greater than that in the trans-azo form at the same pressure. The lateral pressure in the lipid bilayer is normally 30~40 mN/m.[8] Under this pressure, the difference in surface membrane area between the trans and cis form is estimated to be around 11 percent.

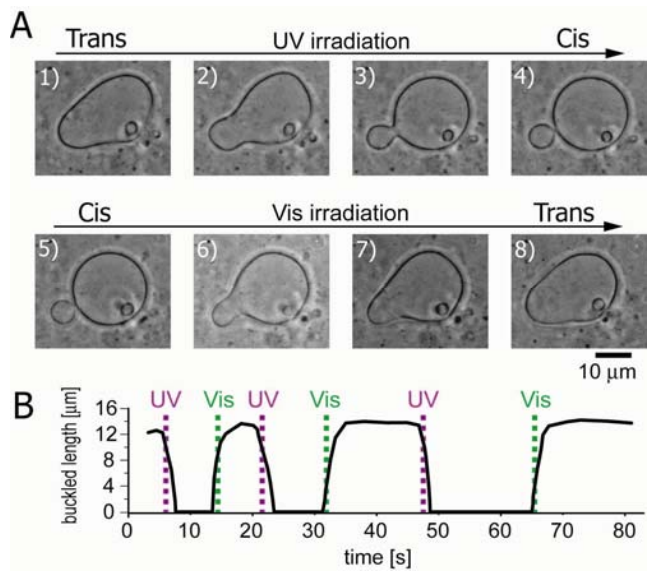

*Figure 2*. A: Photo-induced reversible ellipsoid-bud transition in a cell-sized vesicle. 1) – 4) show the transformation from ellipsoid to bud induced by UV light. The time after starting UV-irradiation is 1) just before light exposure 2) 0.33 s, 3) 1.3 s, and 4) 1.7 s. 5) – 8) show the reverse process from bud to ellipsoid induced by irradiation with green light. The time after green-illumination is 5) just before irradiation, 6) 0.27 s, 7) 0.80 s, and 8) 3.9 s. B: Repetitive photo-switching of the morphology. The time development of the neck length on the buckled part in ellipsoid-bud transformation is shown.

Next, we discuss the mechanism of morphological switching in vesicles. The measurement of Π-A confirmed that the effective cross-section of the photosensitive molecule changes between the trans and cis-azo forms. With reference to a report on water permeability through a phospholipids bilayer membrane[9], the change in volume during transformation over several seconds is estimated to be $\Delta V/V < 10^{-8}$. The volume change is negligible compared to the photo-induced change in surface area on the time-scale in our observation.[10] Thus, the morphological change in Fig. 1 can be attributed to the change in surface area between the isomers. Starting from a spherical vesicle, an increase in surface area causes a large fluctuation without any apparent morphological transition. In such a symmetric vesicle, the membrane area adjusts to the surface area in the spherical water phase. With UV irradiation, the increased area is obliged to exhibit large fluctuation as in Fig. 1A, and the symmetrical shape is preserved without budding. In contrast, an asymmetric vesicle already has enough area for thermal fluctuation, and excess fluctuating area requires much more free energy. The extra area in an asymmetric vesicle is exposed in the budding transition to prevent such an unstable state. The birth of a small bud is induced by an increase in surface area in the encapsulated water phase, since the area of the small budded part is approximately equal to the area of photo-induced expansion [see Supporting Information]. The budded small vesicle is connected to the mother vesicle by a thin phospholipid tube. With visible light, a decrease in the membrane area induces the absorption of the budded vesicle into a single large vesicle. The pathway of the morphological change observed in the present study corresponds to a previous theoretical study on the phase diagram, where the control parameter was temperature instead of a photo-induced change[11]. When the vesicle was some other asymmetrical shape, we also observed several types of transformation: stomatocyte(trans)-prolate(cis) and tubular(trans)-pearling(cis).[12] Again, the initial morphology deformed from the spherical symmetry exhibits a unique shape change. Recently, photoinduced reversible deformations on the assemblies composed of azo compounds have also been reported in the cases of Langmuir-Blodgett film[13], colloids[14] and micelles[15].

In conclusion, we have demonstrated that the photo-isomerization of constituent molecules can switch the morphology when a vesicle exhibits suitable asymmetry. The mechanism of the transformation can be understood in terms of the change in effective cross-sectional area of the photosensitive molecule. The present results may have profound implications and could lead to new ways to control large-scale changes in vesicles using light.

**Acknowledgments.** We thank Prof. S. Nakata, Mr. S. Izuhara, Ms. A. Yamada, Mr. S. Hiromatsu, Mr. H. Okunishi, Mr. H. Kitahata and Dr. M. Takeuchi for their technical assistance in the Langmuir monolayer experiments. T. Hamada is supported by a Research Fellowship from the Japan Society for the Promotion of Science for Young Scientists (No. 16000653).

**Supporting Information Available:** Experimental procedures, Π-A curves of a Langmuir monolayer, the absorption spectra of KAON12 in vesicles and the calculation for the change in surface area during ellipsoid-bud transition.

[†] Kyoto University.
[‡] Osaka City University.
[*] Present address: Kyushu University, Kyusyu 812-8581, Japan.

(1) Alberts, B.; Johnson, A.; Lewis, J.; Raff, M.; Roberts, K.; Walter, P. *Molecular Biology of the Cell*; Garland Science: New York, 2002.
(2) Shum, P.; Kim, J.-M.; Thompson, D. H. *Adv. Drug Del. Rev.* **2001**, *53*, 273.
(3) (a) Lei, Y.; Hurst, J. K. *Langmuir* **1999**, *15*, 3424. (b) Bisby, R.; Mead, C.; Morgan, C. G. *FEBS Lett.* **1999**, *463*, 165. (c) Kano, K.; Tanaka, Y; Ogawa, T; Shimomura, M; Kunitake, T. *Photochem. Photobiol.* **1981**, *34*, 323.
(4) (a) Hamada, T.; Yoshikawa, K. *Chem. Phys. Lett.* **2004**, *396*, 303. (b) Hotani, H.; Nomura, F.; Suzuki, Y. *Curr. Opin. Colloid Interface Sci.* **1999**, *4*, 358. (c) Lipowsky, R. *Nature* **1991**, *349*, 475.
(5) (a) Nomura, S-i. M.; Tsumoto, K.; Hamada, T.; Akiyoshi, K.; Nakatani Y.; Yoshikawa, K. *ChemBioChem* **2003**, *4*, 1172. (b) Luisi, P. L.; Walde, P. *Giant Vesicles*; John Wiley & Sons: Chichester, 2000.
(6) (a) Girard, P.; Prost, J.; Bassereau, P. *Phys. Rev. Lett.* **2005**, *94*, 088102. (b) Manneville, J.-B.; Bassereau, P.; Ramaswamy, S.; Prost, J. *Phys. Rev. E* **2001**, *64*, 021908.
(7) Nagasaki, T.; Taniguchi, A.; Tamagaki, S. *Chem. Lett.* **2003**, *32*, 88.
(8) Adam, N. K. *The Physics and Chemistry of Surfaces*; Clarendon Press: Oxford, 1938.
(9) Fettiplace, R. *Biochim. Biophys. Acta* **1978**, *513*, 1.
(10) Under UV irradiation, the conformation of the cis-azo vesicles gradually shifts to a sphere over several tens of minutes, which is $10^3$ longer than the time of the photo-induced transformation in Fig. 1 and Fig. 2. The membrane composed of the cis-isomer may have higher ion permeability than that of the trans-isomer because of its bulky structure as reported eleswhere[2]. A cis-azo vesicle can easily transfer water molecules across the membrane and soon reaches an equilibrium morphology, i.e. spherical shape with the lowest bending energy.
(11) (a) Miao, L; Seifert, U.; Wortis, M.; Döbereiner H. G. *Phys. Rev. E* **1994**, *49*, 5389. (b) Seifert, U.; Berndl, K.; Lipowsky, R. *Phys. Rev. A* **1991**, *44*, 1182. (c) Kas, J.; Sackmann, E. *Biophys. J.* **1991**, *60*, 825.
(12) (a) Allain, J.-M.; Storm, C.; Roux, A.; Amar, M. Ben; Joanny, J.-F. *Phys. Rev. Lett.* **2004**, *93*, 158104. (b) Tanaka, T.; Tamba, Y.; Masum, S. Md.; Yamashita, Y.; Yamazaki, M. *Biochim. Biophys. Acta* **2002**, *1564*, 173. (c) Brückner, E.; Sonntag, P.; Rehage, H.; *Langmuir*, **2001**, *17*, 2308. (d) Petrov, P. G.; Lee J. B.; Döbereiner H. G., *Europhys. Lett.* **1999**, *48*, 435.
(13) Matsumoto, M.; Miyazaki, D.; Tanaka, M.; Azumi, R.; Manda, E.; Kondo, Y.; Yoshino, N.; Tachibana, H. *J. Am. Chem. Soc.* **1998**, *120*, 1479.
(14) Li, Y.; He, Y.; Tong, X.; Wang X. *J. Am. Chem. Soc.* **2005**, *127*, 2402.
(15) Orihatra, Y.; Matsumura, A.; Saito, Y.; Ogawa, N.; Saji, T.; Yamaguchi, A.; Sakai, H.; Abe, M. *Langmuir* **2001**, *17*, 6072.

**Supporting information**

**Experimental procedures**

        Cell-sized vesicles were prepared from dioleoyl-phosphatidylcholine (DOPC) (Wako, Japan) and a synthesized photosensitive lipid (KAON12) through natural swelling[1] or electroformation[2], where the lipid film was swollen with distilled water ([KAON12]/[DOPC] = 60 μM/100 μM). No apparent difference in the photo-induced effect was noted between these two different methods of preparation. Ten μL of the vesicle solution was placed on a glass slip, covered with another smaller slip at a spacing of 50 μm, and sealed. We observed the vesicles with a phase-contrast microscope (Nikon TE-300, Japan), and irradiated them through a dichroic mirror unit, UV (365 nm) and green (546 nm), with an extra-high pressure mercury lamp (100 W) for photo-isomerization. The images were recorded on digital videotape at 30 frames/s.

*References*

(1) Bangham, A. D.; Standish, M. M.; Watkins, J. C. *J Mol. Biol.* **1965**, *13*, 238.
(2) Dimitrov, D. S.; Angelova, M. I. *Prog. Colloid & Polymer Sci.* **1987**, *73*, 48.

**Langmuir monolayer**

Lipids were dissolved in chloroform/benzene=1/9 to obtain a [KAON12]/[DOPC] = 3/5 solution, which was the same as the ratio in the microscopic observation. A drop of the solution was spread onto pure water contained in a Teflon trough (Kyowa, Japan), and Π-A curves were measured with each isomer (Fig. SI-1). Isomerization into the cis conformer was achieved by irradiating the lipid-dissolved organic solvent in a test tube with a UV illuminator (365 nm, 8 W: UVP, USA) for 15 min.

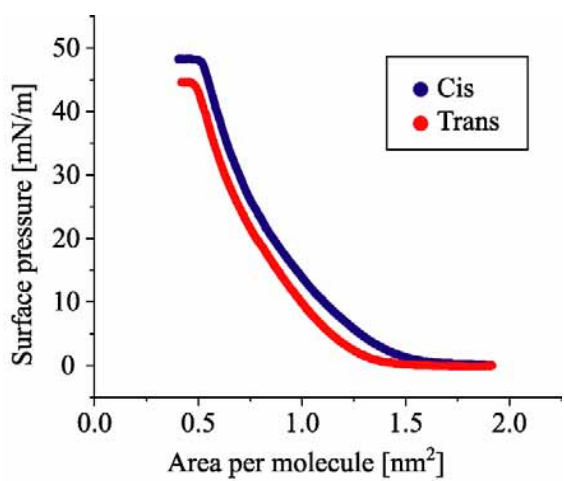

*Figure SI-1.* Π-A curves of a Langmuir monolayer in trans and cis-azo conditions.

**Absorbance**

Absorption spectra of vesicles were measured using a spectrometer U-3210 (Hitachi, Japan) at 24 °C. A vesicle solution ([KAON12]/[DOPC] = 30 µM/50 µM) was added to a two-plane transparent quartz cell, and photo-isomerization was caused by direct irradiation through a dichroic mirror unit, UV (365 nm) and green (546 nm), of a fluorescent microscope (Olympus IX-70, Japan) with an extra-high pressure mercury lamp (100 W).

Figure SI-2 shows the change in the absorption spectra in vesicles upon photo-irradiation. We irradiated all trans-azo vesicles with UV light at spectrum (a) for 30 min, and then the spectrum shifted to (b), where trans/cis = 17/83. Furthermore, 30 min of green-illumination restored the ratio to (c) trans/cis = 79/21. Afterward, this cycle was repeated one more time at (d) trans/cis = 17/83 after UV and (e) trans/cis = 80/20 after green. This confirms that it is possible to reversibly control the structure of the photosensitive molecule inside vesicles by light.

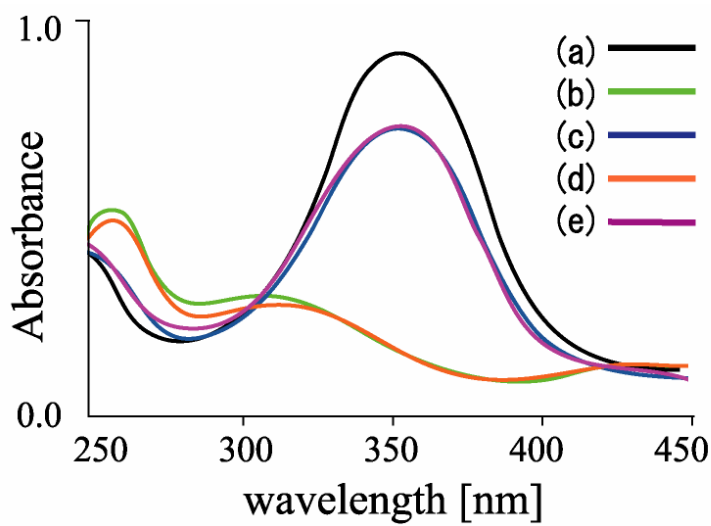

*Figure SI-2*. Absorption spectra of KAON12. The spectrum (a) in an all-trans vesicle was shifted to (b) after 30 min of UV-irradiation. Thirty min of green-illumination gave (c), 30 min of UV-illumination led to (d), and 30 additional min of green-illumination gave (e). The ratio of trans/cis is estimated to be (a) 100/0, (b) 17/83, (c) 79/21, (d) 17/83, and (e) 80/20.

**Change in surface area during the budding transition**

We will analyze the change in surface area in the ellipsoid-bud transition in Fig. 2A. For simplicity, we assume the shape of an ellipsoid vesicle and a spherical vesicle with a bud as in Fig. SI-3, by defining the variables *r*, *R* and *l*. The length *l* is obtained under the condition of volume conservation.

$$l = \frac{2(R^3 - 6r^3)}{R^2 + 2Rr + 4r^2}. \qquad (1)$$

The difference in surface area between the ellipsoid and bud shapes is expressed as

$$\Delta S = S_{budding} - S_{ellipsoid} = 4\pi r^2 \frac{1 + \frac{r}{R} + 2\left(\frac{r}{R}\right)^2}{1 + \frac{r}{R} + \left(\frac{r}{R}\right)^2}. \qquad (2)$$

If $R \gg r$, equation (2) leads to $\Delta S \cong 4\pi r^2$. Thus, the ratio of the surface area shift can be represented as

$$\frac{\Delta S}{S} \cong \frac{r^2}{R^2}. \qquad (3)$$

Substituting $r/R \sim 1/3$ from the microscopic observations in Fig. 2A into equation (3), the photo-induced change in surface area is estimated to be ~ 0.1, which corresponds well to the result in the Π-A measurements (Fig. SI-1).

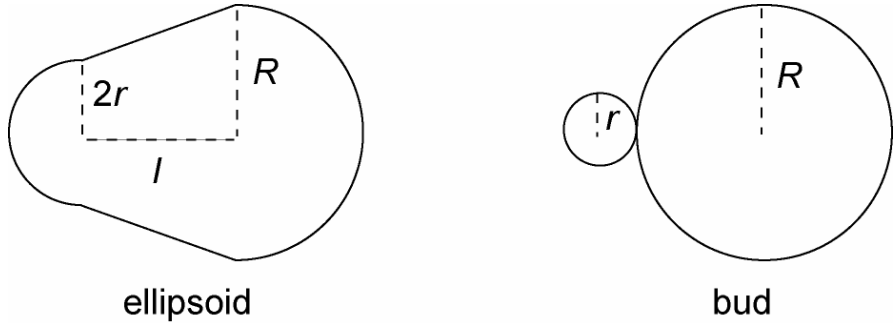

*Figure SI-3* Schematic representation of ellipsoid and budding spherical vesicles, corresponding to the experimental observation in Fig. 2A.

S4